%% file: main.tex
\documentclass[sigconf]{acmart}
\AtBeginDocument{%
  }

\copyrightyear{2025}
\acmYear{2025}
\setcopyright{cc}
\setcctype{by}
\acmConference[CIKM '25]{Proceedings of the 34th ACM International
Conference on Information and Knowledge Management}{November 10--14,
2025}{Seoul, Republic of Korea}
\acmBooktitle{Proceedings of the 34th ACM International Conference on
Information and Knowledge Management (CIKM '25), November 10--14, 2025,
Seoul, Republic of Korea}\acmDOI{10.1145/3746252.3760960}
\acmISBN{979-8-4007-2040-6/2025/11}

\usepackage{threeparttable}
\usepackage{arydshln}
\usepackage{xspace}
\usepackage{graphicx}
\usepackage{listings}
\usepackage{hyphenat}
\usepackage{balance}

\newcommand{\ours}{Promptodile\xspace} 
\begin{document}

\title{Study on LLMs for Promptagator\hyp{}Style Dense Retriever Training}

\author{Daniel Gwon}
\orcid{0009-0002-1283-4769}
\authornote{Both authors contributed equally to this research}
\affiliation{%
    \institution{Massachusetts Institute of Technology Lincoln Laboratory}
    \city{Lexington}
    \state{MA}
    \country{USA}
}
\email{daniel.gwon@ll.mit.edu}

\author{Nour Jedidi}
\orcid{0009-0007-0189-9678}
\authornotemark[1]
\affiliation{
    \institution{University of Waterloo}
    \city{Waterloo}
    \state{ON}
    \country{Canada}}
  \email{njedidi@uwaterloo.ca}
\additionalaffiliation{
    \institution{Massachusetts Institute of Technology Lincoln Laboratory}
    \city{Lexington}
    \state{MA}
    \country{USA}
}

\author{Jimmy Lin}
\orcid{0000-0002-0661-7189}
\affiliation{
    \institution{University of Waterloo}
    \city{Waterloo}
    \state{ON}
    \country{Canada}
}
\email{jimmylin@uwaterloo.ca}

\begin{abstract}
\input{sections/0_abstract}
\end{abstract}



\maketitle
\footnotetext{DISTRIBUTION STATEMENT A. Approved for public release. Distribution is unlimited. This material is based upon work supported by the Department of the Air Force under Air Force Contract No. FA8702-15-D-0001 or FA8702-25-D-B002. Any opinions, findings, conclusions or recommendations expressed in this material are those of the author(s) and do not necessarily reflect the views of the Department of the Air Force.}
\input{sections/1_introduction}
\input{sections/2_background}

\input{sections/3_experiments}
\input{sections/4_analysis}
\input{sections/5_conclusion}
\input{sections/genai_disclosure}
\bibliographystyle{ACM-Reference-Format}
\balance
\bibliography{main}

\end{document}

%% file: sections/0_abstract.tex
Promptagator~\cite{daipromptagator}  demonstrated that Large Language Models (LLMs) with few-shot prompts can be used as task-specific query generators for fine-tuning domain-specialized dense retrieval models. However, the original Promptagator approach relied on proprietary and large-scale LLMs which users may not have access to or may be prohibited from using with sensitive data. In this work, we study the impact of open-source LLMs at accessible scales ($\leq$14B parameters) as an alternative. Our results demonstrate that open-source LLMs as small as 3B parameters can serve as effective Promptagator-style query generators. We hope our work will inform practitioners with reliable alternatives for synthetic data generation and give insights to maximize fine-tuning results for domain-specific applications.\footnote{Our code is available at \url{https://www.github.com/mit-ll/promptodile}}

%% file: sections/1_introduction.tex
\section{Introduction}

Dense retrieval models have greatly enhanced modern Information Retrieval (IR) systems~\cite{karpukhin-etal-2020-dense}, endowing search systems with semantic search capabilities while minimizing the latency associated with models using cross-encoder architectures~\cite{nogueira2019passage}. However, for domains and tasks that lack large amounts of training data, dense retrievers have been shown to struggle, underperforming sparse bag-of-words techniques like BM25~\cite{INR-019}. 

To overcome this challenge, researchers have looked towards employing Large Language Models (LLMs) as dataset generators~\cite{bonifacio2022inpars, daipromptagator}. These techniques typically prompt LLMs to generate synthetic queries that align with given documents from the corpus for training dense retrieval models. A popular approach is Promptagator~\cite{daipromptagator}, which leverages an LLM to amplify a few human-annotated examples from the target domain into many training examples. The core idea behind Promptagator is that different search tasks and domains have different relevance definitions that should be considered when generating synthetic training queries. 

While Promptagator was shown to be highly effective in domain-specific settings, its results were based on a proprietary and large-scale LLM, which users may not have access to or may be prohibited from using with sensitive data. With advancements in open-source LLMs —  particularly at accessible scales ($\leq$14B parameters) — we investigate \emph{\ours}, an open-source variant of Promptagator focusing on the utility of LLMs in three settings: (1) low-resource domains where training data doesn't exist, (2) domains where information needs are found in sensitive data (e.g., PII or proprietary data) that users can't share with popular LLM providers (e.g. OpenAI and Anthropic), and (3) compute-constrained environments where users may not have resources to self-host larger models. 

Towards this goal, we study \ours with 10 open-source LLMs across four LLM families, ranging in scale from 1B-14B parameters, evaluated on seven low-resource BEIR datasets. Our findings can be summarized as follows: (1) Across all LLMs, \ours is generally competitive with Promptagator, demonstrating that recent, accessible LLMs can serve as effective query generators. We highlight that these results came \emph{without} round-trip filtering, suggesting that as LLMs improve, filtering may be less important. (2) Furthermore, we find that smaller LLMs ($\approx$3B parameters) are just as effective as larger ones (between 7B to 14B parameters), demonstrating that there is no benefit in leveraging expensive LLMs for query generation.

We emphasize that we are not the first to study open-source LLMs as query generators~\cite{jeronymo2023inpars, boytsov2023inpars}. Our research builds upon this line of work by focusing on dense retrievers and studying across a wider range of LLMs.  The primary contribution of \ours is to provide insights for practitioners building Promptagator-style retrieval systems in low resource domains amidst compute and data privacy constraints; we do \emph{not} claim improvements on Promptagator. We hope our findings will inform users on how to best maximize dense retriever effectiveness for domain-specific applications. 

%% file: sections/2_background.tex
\section{Promptagator}

In the Promptagator setup, it is assumed that a large collection of $n$ texts, denoted by $\mathcal{D}=\{d_1, d_2, \ldots, d_n\}$, is available but there are a limited number of annotated query-document pairs for training the dense retrieval model. To overcome this challenge, Promptagator proposes to leverage an LLM's few-shot capabilities to amplify a \emph{few} (e.g., 2-8) task-specific query-document pairs into a larger synthetic training dataset.  A critical novelty of Promptagator, compared to similar work like InPars~\cite{bonifacio2022inpars}, is that it recognizes that retrieval tasks have different search intents (i.e., definitions of what relevance means). Thus, when generating domain-specific synthetic training data for a dense retriever, these search intents should be captured in the form of \emph{few-shot prompting}.

The Promptagator workflow is as follows: Given $d_{i}$, a document in $\mathcal{D}$, an LLM is prompted to perform document-to-query generation, conditioned on \textit{annotated}, task-specific query-document pairs, $\{(q_1, d_1), (q_2, d_2)\ldots(q_i, d_i)\}$, where $q_i$ denotes an annotated query for document $d_{i}$. This process is repeated for $k$ documents, where $k \leq n$, in turn generating many synthetic query-document pairs: $\{(\hat{q}_1, d_1), (\hat{q}_2, d_2) \ldots, (\hat{q}_k, d_k)\}$, where $\hat{q}$ denotes a synthetic query generated by the LLM.  These synthetic query-document pairs are subsequently used to train a dense retrieval model.

While Promptagator demonstrated strong performance across various low-resource benchmarks, its results were based on a proprietary 137B FLAN-T5~\cite{chung2024scaling} model, leaving open the question of whether smaller, accessible LLMs can serve as viable alternatives. In this paper, we study the effect of the latest open-source LLMs on performance in Promptagator-style dense retrieval model training.

%% file: sections/3_experiments.tex
\begin{table*}[t!]
  \centering
    \caption{Main results (NDCG@10) across seven BEIR tasks. Note that for our experiments, we do \emph{not} perform round-trip filtering to enable fair comparison across LLMs. As such, we also report Promptagator results without filtering, Promptagator (no filtering), based on Figure 2a in Dai et al.~\cite{daipromptagator}. $^*$ denotes datasets in which \ours uses 10$x$ less data than Promptagator.} 
    \begin{tabular}{l|ccccccc|c}
      \toprule
      \textbf{Retriever} & ArguAna & $\text{DBPedia}^*$ & FiQA & $\text{HotpotQA}^*$ & NFCorpus & SCIDOCS & SciFact & \textbf{Avg.} \\
      \midrule
      BM25 & 39.7 & 31.8 & 23.6 & 63.3 & 32.2 & 14.9 & 67.9 & 39.1  \\
      \hline
      Contriever & 37.9 & 29.2 & 24.5 & 48.1 & 31.7 & 14.9 & 64.9 &  35.9  \\
      Contriever MS-MARCO & 44.6 & 41.3 & 32.9 & 63.8 & 32.8 & 16.5 & 67.7 & 42.8   \\
      Promptagator (with filtering) & 59.4 & 38.0 & \bf 46.2 & 61.4 & 33.4 & \bf 18.4 & 65.0 &  46.0 \\
      Promptagator (no filtering) & 58.2 & 33.2 & 46.3 & 57.6 & 34.1 & 17.1 & 66.4  & 44.7\\
       \hline
      \ours (no filtering) & & & & & & & & \\
      \quad QGen: Llama-3.2-3B  & 57.1 &  37.7  & 36.3 & 62.7  & 34.5 & 17.7 & \bf 69.9 & 45.1  \\
      \quad QGen: Llama-3.1-8B  & 59.4 &  35.4 & 37.7 & 62.1 & 34.1 & 16.7 & 69.9 & 45.0  \\
      \cdashline{1-9}[0.8pt/2pt] 
      \quad QGen: Qwen2.5-3B   & 49.1 &  38.4  & 37.2 & 64.2 & 32.4 & 17.6 & 67.3 & 43.7  \\
      \quad QGen: Qwen2.5-7B  & 59.8 &  37.3 & 35.9 & 64.1 & 33.4 & 16.1 & 64.7 & 44.5  \\
      \quad QGen: Qwen2.5-14B  & 53.7 &  \bf 40.5 & 39.9 & 64.7 & 33.0 & 15.4 & 66.6 & 44.8  \\
      \cdashline{1-9}[0.8pt/2pt]
      \quad QGen: Phi-3-mini  & 56.1 &  37.6 & 36.5 & 62.6 & 34.7 & 18.2 & 69.6 & 45.0  \\
      \quad QGen: Phi-3-medium  & 57.6 &  37.3 & 42.4 & \bf 65.0 & \bf 35.2 & 17.2 & 68.8 & \bf 46.2  \\
      \cdashline{1-9}[0.8pt/2pt]
      \quad QGen: gemma-3-1b  & 54.4 &  31.2 & 20.9 & 59.0 & 32.9 & 12.1 & 66.4 &  39.6  \\
      \quad QGen: gemma-3-4b  & \bf 61.8 &  34.9 & 34.6 & 63.9 & 33.4 & 15.2 & 68.8 & 44.7  \\
      \quad QGen: gemma-3-12b & 58.7 &  35.1 & 37.6 & 63.1 & 33.9 & 11.7 & 65.8 & 43.7 \\
      \bottomrule
    \end{tabular}
  \label{tab:BEIR}
\end{table*}

\begin{table}[t!]
  \centering
   \caption{Impact of backbone dense retriever on NDCG@10. \ours is trained with  Phi-3-medium Qgen model.}
  \resizebox{\columnwidth}{!}{%
  \begin{tabular}{l|cccc|c}
    \toprule
    \bf Retriever & ArguAna & NFCorpus & SCIDOCS & SciFact & \bf Avg. \\
    \midrule
    Contriever & 37.9 & 31.7 & 14.9 & 64.9 & 37.4 \\
    Contriever MS-MARCO & 44.6 & 32.8 & 16.5 & 67.7 &  40.4 \\
    \cdashline{1-5}[0.8pt/2pt] 
    E5 Base (Unsupervised)  & 42.2 & 35.8 & 21.1 & 73.7 & 43.2\\
    E5 Base (Supervised) & 51.4 & 36.6 & 19.0 & 73.1 & 45.0 \\
    \cdashline{1-5}[0.8pt/2pt] 
    Promptagator & 59.4 & 33.4 & 18.4 & 65.0 & 44.1\\
    Promptagator (no filtering) & 58.2  & 34.1 & 17.1 & 66.4 & 44.0 \\
    \hline
    \ours \\
    \quad w/ Contriever &  57.6 &  35.2 &  17.2 & 68.8 & 44.7 \\
    \quad w/ E5 Base (Unsupervised) &  57.5 & \bf 37.6 & \bf 22.5 & \bf 74.9 & \bf 48.1 \\
    \quad w/ Contriever MS-MARCO &  \bf 59.4 &  34.9 &  17.7 &  70.5 & 45.6 \\
    \bottomrule
  \end{tabular} }
  \label{tab:encoder}
\end{table}

\section{Experimental Setup}

We investigated various aspects of synthetic query generation and retrieval model training using select datasets from the BEIR benchmark~\cite{thakur2021beir} to train and evaluate our models. We used vLLM~\cite{woosuk2023vllm} for offline inference on various NVIDIA accelerators.

The selected datasets are detailed in Table \ref{tab:BEIR}. These were chosen to maintain a diversity of tasks, domains, and sizes while managing limited compute. For each dataset, we randomly sampled up to 100K documents to generate synthetic queries; we used the full corpus for datasets smaller than 100K documents.

Prompts were task-specific with few-shot examples, similar to those in Promptagator, and formatted in chat templates. This is an example system, user, and assistant prompt for SciFact:

\begin{lstlisting}
System: You are a high-quality synthetic data generator. Your task is to read an abstract from scientific research literature and generate a relevant scientific claim. A claim is relevant if the abstract contains all of the necessary evidence to support the claim. Use the following examples to guide you. Respond with only the scientific claim.
User: Abstract: {}
Assistant: {}
\end{lstlisting}
We selected few-shot examples from the development split when available, otherwise we used the test split (ArguAna, SCIDOCS, and SciFact only). To ensure our examples were relevant, we sampled from relevant queries with the highest scores. For instance, DBPedia~\cite{hasibi2017dbpedia} provides relevance assessment scores of 0 (irrelevant), 1 (relevant), and 2 (highly relevant); our few-shot examples all had a relevance assessment score of 2. For each example, the document text was injected into the content for the user role and the query text was injected into the content for the assistant role. We did not use document titles and removed the selected examples from the set of documents used to generate synthetic queries.\footnote{Since we removed all few-shot examples from the document sets, we did not remove them when evaluating as they are never exposed to the dense retrieval model.} We prompted an LLM with a task-specific system prompt with few-shot examples and a single document from which to generate a relevant query, and did so repeatedly for all datasets.

We also investigated two sets of sampling parameters (limited to those available in the vLLM library~\cite{woosuk2023vllm}). The first set came from Promptagator, which used a temperature of 0.7 across all models and model defaults elsewhere. The second set was selected to induce more diverse outputs and they are listed in Table \ref{tab:samplingparams}. For all datasets, we generated a maximum of 256 tokens with 8 return sequences, meaning we generate up to 800K training queries for a given combination of LLM, sampling parameters, and dataset.

This study used open-source, transformer-based~\cite{NIPS2017_3f5ee243} decoder-only LLMs with varying model architectures and parameter sizes to generate synthetic queries. We refer to them as \emph{QGen}. See Table \ref{tab:BEIR} for the full list.

For training the dense retrieval model, we utilized the unsupervised Contriever~\cite{izacard2022unsupervised} as the default backbone model. To train Contriever, we followed the setup from Dai et al.~\cite{daipromptagator}, using cross-entropy loss over in-batch random negatives. We trained with a learning rate of 2e-5 and batch size of 128 for datasets with less than 50K documents and a batch size 6000 otherwise. Prior to training, we randomly sampled 10\% of our corpus as an evaluation set. Dense retrieval models were then trained for up to 30 epochs and the checkpoint with the lowest loss over the evaluation set was selected. We employed early stopping if the evaluation loss did not decrease for three consecutive epochs. Training was done using using GradCache~\cite{gao2021scaling} via Sentence Transformers~\cite{reimers-gurevych-2019-sentence}  and retrieval experiments were performed with Pyserini~\cite{lin2021pyserini}. We chose not to perform round-trip filtering as is done in Promptagator to better study the impact of the various QGen models.

\section{Results}

In this section we investigate the retrieval accuracy of \ours across different QGens and datasets. Results are shown in Table~\ref{tab:BEIR}.  

\subsection{\ours versus Promptagator}
When comparing  Promptagator (no filtering) to \ours, we find that, across QGen models, \ours is generally on-par with Promptagator, even when using LLMs as small as 3B parameters. Besides \ours with gemma-3-1b, \ours always scores within one point — scoring between 43.7 to 46.2 NDCG@10 — of Promptagator (no filtering), which achieves an average of 44.7 NDCG@10.  Additionally, we find that \ours with Phi-3-medium performs competitively with Promptagator (\emph{with} filtering), suggesting that as LLMs improve, the filtering step may not be as important. For example, on DBPedia and HotpotQA — datasets where Promptagator gets the largest boost from round-trip filtering — the best \ours models (using Qwen2.5-14B and Phi-3-medium) score 2.5 and 3.4 points higher, respectively, than Promptagator.

\subsection{Impact of LLM Scale on \ours}
When comparing \ours within LLM families, we do not see strong improvements to \ours performance with QGens above 3B parameters. While gemma-3-1b shows the worst performance, scoring an average 39.6 versus the other Qgens which are between 43.7-46.2, we find that the improvement from using a 3B QGen versus a 14B QGen is minimal. In fact, within LLM families, on average \ours with 3B/4B parameter QGens are always within approximately one point of \ours with QGens of 7B+ parameters. These results demonstrate that smaller QGen models can effectively produce Promptagator-style dense retrieval systems, and there is no clear benefit to using larger, more expensive LLMs. The effectiveness of small QGen models makes fine-tuned dense retrieval models highly accessible.

To provide a cost comparison, we count the number of input and output tokens using one HotpotQA synthetic query dataset generated with Llama-3.2-3B as the QGen model\footnote{Generated on 4xA100 GPUs (80GB VRAM) in $\approx$9 hours}. In this dataset, we have $\approx$6.5M input tokens and $\approx$13.8M output tokens. At the time of writing, OpenAI charges \$2/1M input tokens and \$8/1M output tokens giving a total cost of \$124\footnote{We ignore the cost of few-shot prompts which can be cached} to generate synthetic queries for training. While not prohibitive on its own, this cost can grow quickly when factoring in many dozens of training datasets over time. In contrast, Llama-3.2-3B can easily be self-hosted.

\subsection{\ours versus MS-MARCO Transfer}

We compare \ours to Contriever MS-MARCO, and find that under the same backbone dense retrieval model (Contriever), Promptagator\hyp{}style training with small, open-source LLMs is more effective than solely training on a large-scale dataset like MS-MARCO and \emph{transferring} it to a new domain. This result further demonstrates the strong role LLMs can play as task-specific dataset generators. As we show in the next subsection, \ours and large-scale dataset training are not \emph{independent} of each other, and can in fact be complementary. 

\subsection{\ours with Different Retrievers}


Table \ref{tab:BEIR} focused on Contriever as the backbone dense retriever for \ours as it shares similarities to the dense retriever used in the original Promptagator implementation. Here, we study how \ours can benefit using (1) an improved unsupervised dense retriever and (2) a dense retriever first trained on a large-scale dataset like MS-MARCO. To align with our goal of developing a solution for compute-constrained environments, we focus on BERT-base~\cite{devlin-etal-2019-bert} sized models and leave exploration of larger dense retrievers as future work. 

The results for this experiment are shown in Table \ref{tab:encoder}. We find that fine-tuning \ours with E5 Base (Unsupervised)~\cite{wang2022text}, a more effective unsupervised dense retriever than Contriever, can make strong improvements to the performance of \ours. Specifically, fine-tuning from E5 Base (Unsupervised) makes a 3.4 point boost, on average, upon fine-tuning from Contriever across the four datasets.  When leveraging the same base dense retriever, but first fine-tuning on MS MARCO, \ours is able to improve by approximately 1 point on average, from 44.7 to 45.6 NDCG@10.

%% file: sections/4_analysis.tex
\section{Synthetic Query Analysis}

\begin{table}[t!]
    \centering
    \caption{Sampling parameters for more diverse queries.}
    \label{tab:samplingparams}
    \resizebox{\columnwidth}{!}{%
        \begin{tabular}{c c c c c c} \toprule
            temperature & $\text{top}_{p}$ & $\text{top}_{k}$ & repetition\_penalty & presence\_penalty \\ \midrule
            1.0 & 0.95 & 70 & 1.2 & 0.3 \\ \bottomrule
        \end{tabular}
    }
\end{table}

\begin{figure}[t!]
    \centering
    \includegraphics[width=\linewidth]{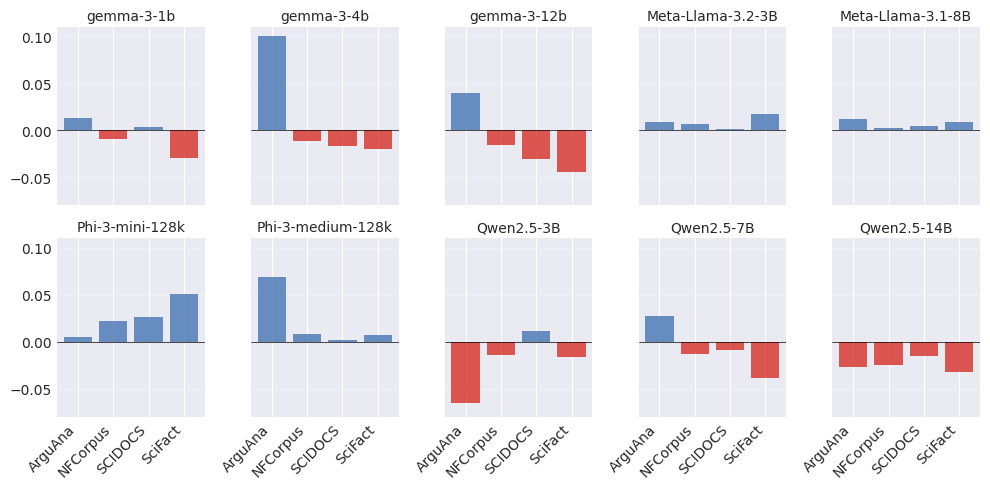}
    \caption{Difference in performance by sampling parameters, ($\text{NDCG@10}_{\text{default}}-\text{NDCG@10}_{\text{creative}}$)}
    \Description{Difference in NDCG@10 of default sampling parameters and creative sampling parameters, organized by model and dataset.}
    \label{figure:samplingparam}
\end{figure}

Dataset diversity is a key component in training performant LLMs~\cite{zhou2023lima,grattafiori2024llama}, and we hypothesize that generating more diverse or creative synthetic queries would also improve dense retrieval performance. We induce our QGen models to generate more creative outputs by changing sampling parameters for four datasets and our results are presented in Figure \ref{figure:samplingparam} with new parameters in Table \ref{tab:samplingparams}. Generally, the impact is model-specific with Gemma and Qwen showing performance gains on more creative sampling parameters whereas Llama and Phi do better on default parameters.  We see large swings in performance on ArguAna, highlighting the variation in performance one might expect on challenging tasks.

\begin{table}[t!]
    \centering
    \caption{Relationship between verbosity and performance (using Spearman’s rank correlation of NDCG@10 and normalized words per query).}
    \label{tab:verbosity}
    \begin{tabular}{l r r c} \toprule
        dataset & $\rho$ & p-value & n \\ \midrule
        FiQA & 0.5758 & 0.0816 & 10 \\
        NFCorpus & 0.6727 & 0.0233 & 11 \\
        SCIDOCS & 0.6848 & 0.0289 & 10 \\ \bottomrule
    \end{tabular}
\end{table}

\begin{figure}[t!]
    \centering
    \includegraphics[width=\linewidth]{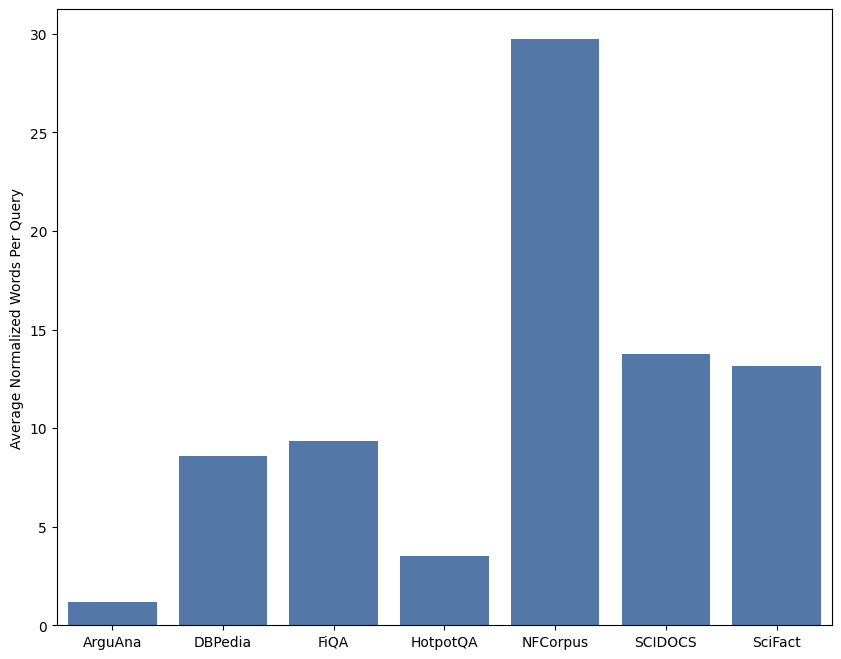}
    \caption{Verbosity by task.}
    \Description{Vertical bar chart showing average normalized words per query by dataset.}
    \label{fig:verbosity}
\end{figure}

We then consider the importance of verbosity and measure it as the number of words per query, emphasizing words over tokens to facilitate comparison across different models' tokenizers. In this section, we investigate only QGen models with default sampling parameters. To adjust for varying lengths of documents ($d$), we normalize words per query ($q$) for $8$ queries per document in each dataset ($D$) and take the average:
\begin{displaymath}
    \bar{r}_i=\frac{1}{8*|D_i|}\sum_{d \in D_i}\sum_{q \in d}\frac{|d|}{|q|}
\end{displaymath}
where $\bar{r}_i$ is the average normalized length for the $i^{th}$ dataset. Documents are typically longer than queries, which we limit to 256 tokens, so we divide the length of the \emph{document} by the length of the \emph{query} for readability. This means verbosity increases as $\bar{r}_i$ decreases. As we can see in Figure \ref{fig:verbosity}, QGen models adapt well to various retrieval tasks, moderating the length of their queries depending on the given dataset. There is some support that verbosity can impact performance (see Table \ref{tab:verbosity}).

Finally, we use Meta-Llama-3.3-70B-Instruct as a judge~\cite{zheng2023judge} to score queries for ArguAna and NFCorpus on a scale of 0-3 as detailed by Thomas et al.~\cite{thomas2024predict}. Surprisingly, we find no overall correlation between relevance assessment scores (i.e., qrels) and NDCG@10 (standardized by dataset), but see a significant \emph{negative} correlation with default sampling parameters ($\rho$=-0.4818, p-value=0.0270, n=21). We offer two hypotheses: (1) synthetic query-document pairs with higher average qrels have a higher rate of duplication. That is, individual queries are high-quality, but each set of eight queries are similar to one another, biasing average scores upward while overfitting, which is more pronounced on these smaller datasets ($<$10K documents each).\footnote{In contrast, creative sampling parameters using a higher temperature, and therefore considered more diverse, are positively correlated with performance with some confidence ($\rho$=0.3654, p-value=0.1131, n=20)~\cite{wang-2021-GPL}.} (2) Irrelevant synthetic queries are likely to overlap significantly with source documents since the queries are conditioned on those documents, exposing a limitation of LLMs as relevance assessors \cite{alaofi2024llms, clarke2024llmrel}. We leave further investigation to future work.

%% file: sections/5_conclusion.tex
\section{Conclusion}

In this work, we study how small-scale open-source LLMs can serve as query generators for Promptagator-style dense retriever training, which we refer to as \emph{\ours}. Our results demonstrate that users can achieve competitive accuracy using LLMs as small as 3B parameters, making Promptagator-style fine-tuning of dense retrievers accessible for out-of-domain and low-compute settings.

%% file: sections/genai_disclosure.tex
\section*{GenAI Usage Disclosure}

Generative AI tools were solely utilized to assist in making grammatical edits for sections of this work, including the text and tables. 